\documentclass[prb,aps,onecolumn, showpacs]{revtex4}
\usepackage{amssymb}
\usepackage{amsmath}
\usepackage[dvips]{graphicx}
\usepackage[english]{babel}

\begin{document}

\title{Microscopic derivation of Frenkel excitons in second quantization}
\author{Monique Combescot$^{1}$ and Walter Pogosov$^{1,2}$}
\affiliation{$^{1}$Institut des Nanosciences de Paris, Universite Pierre et Marie Curie,
CNRS, Campus Boucicaut, 140 rue de Lourmel, 75015 Paris}
\affiliation{$^{2}$Institute for Theoretical and Applied Electrodynamics, Russian Academy
of Sciences, Izhorskaya 13/19, 125412 Moscow}
\date{\today }

\begin{abstract}
Starting from the microscopic hamiltonian describing free electrons in a
periodic lattice, we derive the hamiltonian appropriate to Frenkel excitons.
This is done through a grouping of terms different from the one leading to
Wannier excitons. This grouping makes appearing the atomic states as a
relevant basis to describe Frenkel excitons in the second quantization.
Using them, we derive the Frenkel exciton creation operators as well as the
commutators which rule these operators and which make the Frenkel excitons
differing from elementary bosons. The main goal of the present paper is to
provide the necessary grounds for future works on Frenkel exciton many-body
effects, with the composite nature of these particles treated exactly
through a procedure similar to the one we have recently developed for
Wannier excitons.
\end{abstract}

\maketitle

\section{Introduction}

The absorption of photons in a dielectric solid can lead to delocalized
excitations called excitons. These excitons are essentially of two types:
the Wannier excitons \cite{Wannier} and the Frenkel excitons \cite{Frenkel}.

Wannier excitons are formed in inorganic semiconductors. The relative motion
of the electron and the hole from which the excitons are made, encompasses
hundred of unit cells. This leads to a small binding energy ($\simeq 10$
meV) and a large Bohr radius ($\simeq 10$ nm). As a consequence, Wannier
excitons start to interact at relatively low densities, giving rise to a
large variety of many-body effects associated with optical nonlinearities,
which makes these conventional semiconductors used in today's technologies.

The second type of exciton, known as Frenkel exciton, is commonly found in
organic crystals. These crystals are of potential importance for future
electronic devices, which makes them under current intensive studies \cite%
{8,9,10,11,12,13}. Frenkel excitons are formed with electron and hole
localized on a small scale, of the order of a single molecular block ($%
\simeq 1$ nm), the typical binding energy being of the order of 1 eV. Due to
interactions between blocks, these single molecule excitations are
transferred from site to site, giving rise to a wave known as Frenkel
exciton.

These pictures can be qualitatively understood by noting that, in
conventional (inorganic) semiconductor, the relative dielectric constant is
rather large ($\simeq 10$), which makes the screening of the interaction
between carriers quite strong. As a result, the attraction between electrons
and holes is weak, which explains the large extension of their relative
motion wave function. On the opposite, the small relative dielectric
constant of molecular crystals ($\simeq 1$) leads to a strong electron-hole
attraction which localizes the pair on a system unit cell.

The separation between Wannier and Frenkel excitons, of course, is not very
sharp. Recent works \cite{10} have shown the limitations of the simple
Frenkel picture for excitons in organic semiconductors and the necessity to
introduce "charge-transfer excitons" in which the electron and the hole are
located in different sites, with similarity to Wannier excitons for which
the distance between electron and hole is large compared to the ion-ion
distance. Along the same line, recent progresses in device nanofabrication
now allow one to obtain organic/inorganic semiconductor structure in which
the hybridization between Frenkel and Wannier excitons can be produced \cite%
{8,9}.

Due to their small exciton Bohr radius, interactions between Frenkel
excitons are expected to occur at a much larger density than the one at
which many-body effects between Wannier excitons start to be noticeable.
This barely comes from the fact that the dimensionless parameter which
controls these many-body effects is%
\begin{equation}
\eta =N(a_{x}/L)^{D}.  \tag{1.1}  \label{1.1}
\end{equation}%
where $a_{x}$\ is the exciton Bohr radius, $L$ the sample size, $D$ the
space dimension, and $N$ the exciton number - the exciton density being $%
n=N/L^{D}$.

Most likely, as for Wannier excitons, interactions between Frenkel excitons
are going to be of importance in electronic devices constructed with organic
semiconductors. This is why a correct treatment of these interactions is
highly desirable. Being made of indistinguishable carriers, Frenkel
excitons, like Wannier excitons, are not well defined objects, which makes
the interactions \textit{between} excitons not possible to identify
properly. As a direct consequence, one cannot describe these interactions
through a potential, as usually done by lack of a correct procedure.

Over the last few years, we have developed a new many-body theory \cite{a,b}
for Wannier excitons, in which the composite nature of the particles is
treated exactly. We have shown that Wannier excitons predominantly interact
through the Pauli exclusion principle which exists between their fermionic
components. This exclusion gives rise to very many carrier exchanges between
excitons which are nicely visualized through Shiva diagrams \cite{M1},
rather different from Feynman diagrams due to the composite character of the
particles. All our works on Wannier excitons end with the same conclusion:
it is not possible to replace composite excitons by elementary excitons, as
commonly done through sophisticated bosonization procedures \cite{M2};
either one misses terms as large as the ones kept, or, in optical nonlinear
effects, one even misses the dominant terms \cite{M3}. This can be readily
seen from a dimensional argument: the Pauli scatterings associated with
carrier exchanges are dimensionless, while the Coulomb scatterings are
energy-like quantities; so that they have to appear with an energy
denominator, which can only be a photon detuning. This makes these Coulomb
terms completely negligible in front of the pure exchange terms - missed
with bosonized excitons - when unabsorbed photons have a large detuning.

The development of a similar procedure for Frenkel excitons requires to
settle a second quantization formalism for these excitons on a clean basis
in order to possibly keep their composite boson nature exactly all over the
calculations. In contrast, the works on interacting Frenkel excitons, we
have up to now seen \cite{Mukamel, Agranovich}, contain a potential written
as $\sum J_{mn}B_{m}^{\dagger }B_{n}$, where $B_{n}^{\dagger }$ creates an
excitation on site $n$. Besides the fact that interactions between excitons
cannot be written as a potential due to the composite nature of these
particles, by writing such a potential in terms of $B_{n}^{\dagger }$ only,
one obviously losses the composite electron-hole nature of this excitation
which exists in the exact potentials between electrons and holes (as seen
below from Eqs. (4.6,13). This is exactly this composite nature that we want
to treat properly, since we know its importance in the case of Wannier
excitons.

In this first work on Frenkel excitons, we propose a microscopic approach to
the description of these excitons based on a second quantization formalism,
starting from the hamiltonian of free electrons in a periodic lattice.
Through a grouping of terms different from the one leading to Wannier
excitons, we introduce the atomic states as a physically relevant
one-electron basis for Frenkel excitons and we rewrite the system
hamiltonian in terms of electrons and holes localized on atomic sites in
order to have a precise description of the interactions. Since we use a
second quantization scheme in terms of electrons and holes separately - and
not in terms of their product $B_{n}^{\dagger }$ as usually done - our
approach automatically takes into account the fermionic composite nature of
the particles forming the Frenkel excitons. This is going to be of crucial
importance for a proper study of many-body effects involving these excitons.
The present preliminary work actually provides the necessary grounds for
further works on Frenkel exciton systems. In a forthcoming publication, we
are going to use this second quantization formalism to derive the Coulomb
and Pauli scatterings of two Frenkel excitons: these are the elementary
scatterings on which all many-body effects dealing with excitons are based.
We will then use these scatterings to calculate the ground state energy of $%
N $ Frenkel excitons in the low density limit, a physical quantity of basic
relevance.

The present paper is organized as follows. In Section II, we start with the
first quantization description of a periodic system made of interacting ions
and electrons. We discuss the conceptual difference between Wannier and
Frenkel excitons which leads to a different grouping of terms in the
hamiltonian. We introduce the atomic states as the appropriate one-electron
basis for the problem when the tight-binding approximation, which neglect
the overlaps between the atomic wave functions for electrons on different
sites, is valid. We also discuss the conceptual difficulty associated to the
atomic basis compared to the free electron and hole basis used in the case
of Wannier excitons. In Section III, we derive the semiconductor hamiltonian
appropriate to Frenkel excitons in second quantization using this atomic
basis. In Section IV, we switch to holes and we reduce the hamiltonian to
the terms which conserve the number of electron-hole pairs. We then discuss
all these terms with a particular attention to the one responsible for the
excitation transfer from site to site. In Section V, we identify the lowest
excited states of the hamiltonian in the absence of interactions between
sites and we show that they form a degenerate subspace. This degeneracy is
split to give rise to Frenkel excitons by the intersite interactions. They
are introduced in Section VI, which is devoted to the precise derivation of
the creation operators for Frenkel exciton and the commutation rules which
govern these operators. The precise handling of these commutators are at the
basis of the many-body theory we are going to construct. We show that, like
Wannier excitons, Frenkel excitons are composite bosons, their commutation
rules differing from the ones of elementary bosons due to the fermionic
nature of the electrons and holes forming these excitons. In Section VII, we
collect the main results of this paper and conclude.

\section{First quantization description}

\subsection{Semiconductor Hamiltonian in first quantization}

Let us consider $N_{s}$ electrons with charge $-\left\vert e\right\vert $,
located at $\mathbf{r}_{i}$ and $N_{s}$ ions with charge $+\left\vert
e\right\vert $, located at $\mathbf{R}_{n}$, with $i$ and $n$ running from 1
to the number of sites $N_{s}$. \ In this first work on Frenkel excitons, we
are going to forget all spin degrees of freedom for the sake of simplicity.
This physically corresponds to have all the electrons with the same spin.
This also means that we drop all degeneracies coming from the orbital part
of the electronic levels. These spin and orbital degrees of freedom generate
very interesting polarization effects. They, however, lead to heavy
notations which are wise to avoid in a first work.

The semiconductor Hamiltonian can be written as%
\begin{equation}
H=H_{kin}+V_{e-ion}+V_{ee}+V_{ion-ion}.  \tag{2.1}  \label{2.1}
\end{equation}%
The one-body operator $H_{kin}$ describes the electron kinetic energy

\begin{equation}
H_{kin}=\sum_{i=1}^{N_{s}}\frac{p_{i}^{2}}{2m},  \tag{2.2}  \label{2.2}
\end{equation}%
where $m$ is the \textit{free} electron mass. The second term of Eq. (2.1),
which describes the electron-ion Coulomb attraction, also is a one-electron
operator. It reads 
\begin{equation}
V_{e-ion}=\sum_{i=1}^{N_{s}}\sum_{n=1}^{N_{s}}\frac{-e^{2}}{\left\vert 
\mathbf{r}_{i}-\mathbf{R}_{n}\right\vert }.  \tag{2.3}  \label{2.3}
\end{equation}

The third operator $V_{ee}$ describes the Coulomb repulsion between
electrons. This two-body operator is given by

\begin{equation}
V_{ee}=\frac{1}{2}\sum_{i\neq j}\sum \frac{e^{2}}{\left\vert \mathbf{r}_{i}-%
\mathbf{r}_{j}\right\vert }.  \tag{2.4}  \label{2.4}
\end{equation}

The last term, $V_{ion-ion}$, which describes the Coulomb interaction
between ions, is a constant with respect to the electron motion. It is
however necessary to keep it in order to work with the hamiltonian of a
fully neutral system. This is required for the convergence of the Coulomb
terms in the large sample limit. We will see below the importance of this
point.

\subsection{Conceptual difference between Wannier and Frenkel excitons}

\subsubsection{Wannier excitons}

Wannier excitons are constructed on delocalized electrons excited from the
valence band to the conduction band. These semiconductor bands result from
the periodic ionic structure of the semiconductor lattice. A simple way to
make these bands appearing is to add and substract a one-electron operator

\begin{equation}
\overline{V}_{ee}=\sum_{i}\overline{v}_{ee}(r_{i}),  \tag{2.5}  \label{2.5}
\end{equation}%
to the semiconductor hamiltonian $H$.\ Yet arbitrary, $\overline{V}_{ee}$
physically represents a mean electron-electron interaction. We will show
below the appropriate way to choose it.

This leads us to rewrite the semiconductor hamiltonian $H$, given by Eq.
(2.1), as

\begin{equation}
H=H_{0}^{(W)}+V_{coul},  \tag{2.6}  \label{2.6}
\end{equation}%
where $V_{coul}$, usually called "semiconductor Coulomb interaction",
corresponds to the difference

\begin{equation*}
V_{coul}=V_{ee}-\overline{V}_{ee},
\end{equation*}%
The zero order hamiltonian for Wannier excitons $H_{0}^{(W)}$ is a sum of
one-electron operators. It can thus be written as

\begin{equation}
H_{0}^{(W)}=H_{kin}+V_{e-ion}+\overline{V}_{ee}+V_{ion-ion}=%
\sum_{i}h_{i}^{(W)},  \tag{2.7}  \label{2.7}
\end{equation}%
where the one-electron operator $h_{i}^{(W)}$ is given by%
\begin{equation}
h_{i}^{(W)}=\frac{p_{i}^{2}}{2m}+\sum_{n}\frac{-e^{2}}{\left\vert \mathbf{r}%
_{i}-\mathbf{R}_{n}\right\vert }+\overline{v}_{ee}(r_{i})+\frac{1}{N_{s}}%
V_{ion-ion}=\frac{p_{i}^{2}}{2m}+v(\mathbf{r}_{i}).  \tag{2.8}  \label{2.8}
\end{equation}%
This one-electron operator has the lattice periodicity if $\overline{v}%
_{ee}(r)$\ is chosen with such a periodicity. Besides this requirement, we
must also enforce $\overline{v}_{ee}(r)$ to be such that the resulting
interaction $v(r)$\ defined in Eq. (2.8) fulfills

\begin{equation}
\int d\mathbf{r}\text{ }v(\mathbf{r})=0,  \tag{2.9}  \label{2.9}
\end{equation}%
in order for $h^{(W)}$ to be the hamiltonian of a fully neutral system. The
simplest choice for $\overline{v}_{ee}(r)$ is to take it as a constant,
through the so-called "positive jellium", namely, $\overline{v}%
_{ee}(r)=N_{s}^{-1}V_{ion-ion}$.

Due to the periodicity of the potential $v(r)$, the eigenstates of $h^{(W)}$
are made of delocalized states separated by band gaps. The relevant ones for
the physics of Wannier excitons belong to the last filled band, called
valence band, and the first empty band, called conduction band. Close to
these band extrema, the eigenstate energies can be written as $\Delta +\hbar
^{2}\mathbf{k}^{2}/2m_{c}$ and $\hbar ^{2}\mathbf{k}^{2}/2m_{v}$, where $%
\Delta $ is the band gap, while $m_{c}$\ and $m_{v}$ are the electron masses
for the conductance and valence bands dressed by the lattice periodic
potential. Note that the effective mass for valence electrons, which are
close to a maximum is negative; so that resulting valence hole mass, defined
as $m_{h}=-m_{v}$, is positive.

The eigenstates $\left\vert \nu k\right\rangle $ of the hamiltonian $h^{(W)}$
with $\nu \ $equal to $v$ or $c$ for valence and conduction states 
\begin{equation}
(h^{(W)}-\varepsilon _{\nu k})\left\vert \nu k\right\rangle =0  \tag{2.10}
\label{2.10}
\end{equation}%
are then used as a one-electron basis, to rewrite the semiconductor
hamiltonian $H$ in the second quantization.

\subsubsection{Frenkel excitons}

The situation for Frenkel excitons is totally different: while Wannier
excitons are constructed on delocalized valence and conduction electron
states, the physical picture of the semiconductor excitations giving rise to
Frenkel excitons is a set of electrons tight to their ions, these electrons
switching from the atomic ground state to the atomic first excited level.
Consequently, the physically relevant one-electron states for Frenkel
excitons are not the free (delocalized) electrons in a periodic lattice,
used in the case of Wannier excitons, but instead the electron localized
atomic states associated to the various ion sites.

In order to make these physically relevant atomic states appearing, we are
led to perform a grouping of terms in the semiconductor hamiltonian $H$,
given in Eq. (2.1), different from the one we have done for Wannier
excitons. This new grouping of terms is%
\begin{equation}
H=H_{0}^{(F)}+V_{e-e}+V_{ion-ion}.  \tag{2.11}  \label{2.11}
\end{equation}%
The zero order hamiltonian for Frenkel excitons $H_{0}^{(F)}$ still is a
one-electron operator, but it now contains the electron kinetic contribution
plus the electron-ion potential; so that it differs from the zero order
hamiltonian for Wannier excitons $H_{0}^{(W)}$. It precisely reads 
\begin{equation}
H_{0}^{(F)}=H_{kin}+V_{e-ion}=\sum_{i}h_{i},  \tag{2.12}  \label{2.12}
\end{equation}%
where $h_{i}$ is now given by 
\begin{equation}
h_{i}=\frac{p_{i}^{2}}{2m}-\sum_{n=1}^{N_{s}}\frac{e^{2}}{\left\vert \mathbf{%
r}_{i}-\mathbf{R}_{n}\right\vert }.  \tag{2.13}  \label{2.13}
\end{equation}

\subsection{Atomic states}

We can note that, in the one-electron hamiltonian $h_{i}$, enter the
interactions of the electron $i$ with \textit{all} the ions $n$; so that $%
h_{i}$ differs from a simple atomic hamiltonian. Nevertheless, it is rather
clear that the physically relevant one-electron states for Frenkel excitons
are going to be these atomic states, i.e., the eigenstates of one electron
in presence of \textit{one} ion. Let us introduce them.

(i) The atomic states $\left\vert \nu \right\rangle $ for one ion located at 
$R=0$, associated to the Hamiltonian

\begin{equation}
h_{atom}=\frac{p^{2}}{2m}-\frac{e^{2}}{r},  \tag{2.14}  \label{2.14}
\end{equation}%
\ are such that

\begin{equation}
(h_{atom}-\varepsilon _{\nu })\left\vert \nu \right\rangle =0,  \tag{2.15}
\label{2.15}
\end{equation}%
their wave functions being $\varphi _{\nu }(\mathbf{r})=\left\langle \mathbf{%
r}|\nu \right\rangle $. As the Hamiltonian eigenstates form an orthogonal
set, we do have

\begin{equation}
\left\langle \nu ^{\prime }|\nu \right\rangle =\int d\mathbf{r}\varphi _{\nu
^{\prime }}^{\ast }(\mathbf{r})\varphi _{\nu }(\mathbf{r})=\delta _{\nu
^{\prime }\nu }.  \tag{2.16}  \label{2.16}
\end{equation}

(ii) If we still consider one ion but located at $R_{n}$ instead of $R=0$,
the corresponding atomic Hamiltonian reads

\begin{equation}
h^{(n)}=\frac{p^{2}}{2m}-\frac{e^{2}}{\left\vert \mathbf{r}-\mathbf{R}%
_{n}\right\vert }.  \tag{2.17}  \label{2.17}
\end{equation}

Due to translational invariance, the eigenstates of $h^{(n)}$ read in terms
of the atomic hamiltonian states ($\varepsilon _{\nu }$, $\left\vert \nu
\right\rangle $) as 
\begin{equation*}
(h^{(n)}-\varepsilon _{\nu })\left\vert \nu n\right\rangle =0,
\end{equation*}%
their wave function being such that

\begin{equation}
\left\langle \mathbf{r}|\nu n\right\rangle =\varphi _{\nu n}(\mathbf{r}%
)=\varphi _{\nu }(\mathbf{r-R}_{n}).  \tag{2.18}  \label{2.18}
\end{equation}

For a given ion $R_{n}$, the bound and extended states of the hamiltonian $%
h^{(n)}$ form a complete basis for one-electron states; so that we do have

\begin{equation}
\left\langle n\nu ^{\prime }|\nu n\right\rangle =\delta _{\nu ^{\prime }\nu
},  \tag{2.19a}  \label{2.19a}
\end{equation}

\begin{equation}
I=\sum_{\nu }\left\vert \nu n\right\rangle \left\langle n\nu \right\vert . 
\tag{2.19b}  \label{2.19b}
\end{equation}%
with the sum restricted to the atomic levels $\nu $.

(iii) If we now turn to the hamiltonian $H_{0}^{(F)}$, given in Eq. (2.13),
we see that it differs from a bare sum of atomic hamiltonians, since each
electron feels the interaction of all the other ions. This is a real
difficulty: unlike Wannier excitons, in which the eigenstates of $%
H_{0}^{(W)} $ can be obtained exactly, the diagonalization of the one-body
part of $H_{0}^{(F)}$ for Frenkel excitons can only be approximated due to
this multiple ion interaction. As shown below, this will force us to make
assumption on the atomic wave functions extensions, the Frenkel exciton
picture being appropriate when the tight-binding approximation is valid.

Due to the sum over all ions contained in the one-body hamiltonian $%
H_{0}^{(F)}$ given in Eq. (2.13), it is, on the one hand, clear that the
states $\left\vert \nu n\right\rangle $, for a \textit{fixed} $n$ (which
form a complete set for one-electron states due to Eq. (2.19)) cannot be a
physically relevant basis to describe Frenkel excitons made of excitations
on all possible ion sites. On the other hand, it is also clear that if we
now leave $n$ running over all the ion positions, the states $\left\vert \nu
n\right\rangle $ for all $\nu $ and all $n$ form an overcomplete set - the
states $\left\vert \nu n\right\rangle $ for one particular $n$ forming a
complete set already. As a bare consequence of this overcompleteness, the
states $\left\vert \nu n\right\rangle $ with different $n$ are not exactly
orthogonal.

However, when the atomic states $\nu $ and $\nu ^{\prime }$ are highly
localized compared to the ion-ion distance, the wave function overlap
between different atoms is essentially zero

\begin{equation}
\varphi _{\nu ^{\prime }}^{\ast }(\mathbf{r-R}_{n^{\prime }})\varphi _{\nu }(%
\mathbf{r-R}_{n})\simeq 0\text{ for }n^{\prime }\neq n,  \tag{2.20}
\label{2.20}
\end{equation}%
whatever $r$ is. Consequently, the scalar product of such atomic states
reduces to

\begin{equation}
\left\langle n^{\prime }\nu ^{\prime }|\nu n\right\rangle =\int d\mathbf{r}%
\varphi _{\nu ^{\prime }}^{\ast }(\mathbf{r-R}_{n^{\prime }})\varphi _{\nu }(%
\mathbf{r-R}_{n})\simeq \delta _{n^{\prime }n}\delta _{\nu ^{\prime }\nu } 
\tag{2.21}  \label{2.21}
\end{equation}%
due to Eq. (2.16): These states $\left\vert \nu n\right\rangle $ are thus
quasi-orthogonal.

(iv) In spite of these difficulties, linked to the overcompleteness of the
states $\left\vert \nu n\right\rangle $ with $n$ running over all ion sites,
the $\left\vert \nu n\right\rangle $\ states have to play a role in the
Frenkel exciton physics. This is why we are going to use them in the second
quantization description of Frenkel excitons.

Before going further, let us add some comments on using these states $%
\left\vert \nu n\right\rangle $ as a basis for second quantization. It is
clear that Eqs. (2.20) is not valid for atomic extended states, nor even for
the highest bound levels. The states $\left\vert \nu n\right\rangle $ can
however be seen as a nice basis for one-electron states if the problem at
hand relies on the highly localized (lowest) atomic states, for which the
overlaps between atomic wave functions for different ions is negligible.
This is actually the case for conventional Frenkel excitons, in which the $%
\nu $'s of interest reduce to the ground state ($\nu =0$) and the first
excited state ($\nu =1$).

A somewhat cleaner way to present the use of these overcomplete states $%
\left\vert \nu n\right\rangle $ is to say that we can always add to the two
sets of atomic states of physical interest, namely, $\left\vert \nu
=0,n\right\rangle $ and $\left\vert \nu =1,n\right\rangle $, other states
constructed "in an appropriate way" , in order to form a complete orthogonal
basis when added to the two states $\left\vert \nu =(0,1),n\right\rangle $.
In problems physically controlled by the two lowest atomic states, these
additional "appropriate states" are not going to play a role in the final
results. This is why it is far simpler not to consider them at all and to
stay with the full overcomplete set of states $\left\vert \nu n\right\rangle 
$ for all $n$ and all $\nu $, the $\nu $'s different from $(0,1)$ playing no
role in the end. As a direct consequence, in the following, the sums over $%
\nu $ will have to be considered as sums over $\nu =(0,1)$.

\section{The Frenkel exciton hamiltonian in the second quantization}

\subsection{One-electron creation operators relevant to Frenkel excitons}

Let $a_{\nu n}^{\dagger }$ be the creation operator for the atomic state $%
\left\vert \nu n\right\rangle $,%
\begin{equation}
\left\vert \nu n\right\rangle =a_{\nu n}^{\dagger }\left\vert \upsilon
\right\rangle .  \tag{3.1}  \label{3.1}
\end{equation}

Since the state $\left\vert \nu n\right\rangle $ expands on the plane wave
basis $\left\vert \mathbf{k}\right\rangle $ as $\left\vert \nu
n\right\rangle =\sum_{\mathbf{k}}\left\vert \mathbf{k}\right\rangle
\left\langle \mathbf{k}|\nu n\right\rangle $, the atomic state creation
operator $a_{\nu n}^{\dagger }$ reads in terms of the free electron creation
operator $a_{\mathbf{k}}^{\dagger }$ as 
\begin{equation}
a_{\nu n}^{\dagger }=\sum_{\mathbf{k}}a_{\mathbf{k}}^{\dagger }\left\langle 
\mathbf{k}|\nu ,n\right\rangle .  \tag{3.2}  \label{3.2}
\end{equation}

By using the fact that the electron operators anticommute, $\left[ a_{%
\mathbf{k}},a_{\mathbf{k}^{\prime }}^{\dagger }\right] _{+}=\delta _{\mathbf{%
kk}^{\prime }}$, it is easy to show, using the above relation, that the
anticommutator for atomic state destruction operators is exactly zero%
\begin{equation}
\left[ a_{\nu ^{\prime }n^{\prime }},a_{\nu n}\right] _{+}=0,  \tag{3.3}
\label{3.3}
\end{equation}%
while for the lowest highly localized states $\nu =(0,1)$, we do have, due
to Eq. (2.21),

\begin{equation*}
\left[ a_{\nu ^{\prime }n^{\prime }},a_{\nu n}^{\dagger }\right] _{+}=\sum_{%
\mathbf{k}}\left\langle \nu ^{\prime }n^{\prime }|\mathbf{k}\right\rangle
\left\langle \mathbf{k}|\nu n\right\rangle =\left\langle \nu ^{\prime
}n^{\prime }|\nu n\right\rangle \simeq \delta _{\nu ^{\prime }\nu }\delta
_{n^{\prime }n}.
\end{equation*}

\subsection{One-body part of the Frenkel exciton hamiltonian}

Let us now use these operators to rewrite the part $H_{0}^{(F)}$\ of the
semiconductor hamiltonian given by Eq. (2.12). Since $H_{0}^{(F)}$\ is a sum
of one-body operators, it can be written in terms of the creation operators
for the one-electron states $\left\vert \nu n\right\rangle $ relevant to
Frenkel excitons, as%
\begin{equation}
H_{0}^{(F)}=\sum_{\substack{ \nu ^{\prime }n^{\prime }  \\ \nu n}}%
\varepsilon (\nu ^{\prime }n^{\prime },\nu n)a_{\nu ^{\prime }n^{\prime
}}^{\dagger }a_{\nu n}.  \tag{3.4}  \label{3.4}
\end{equation}

According to the second quantization procedure, the prefactor $\varepsilon
(\nu ^{\prime }n^{\prime },\nu n)$ is given by%
\begin{equation}
\varepsilon (\nu ^{\prime }n^{\prime },\nu n)=\int d\mathbf{r}\text{ }%
\varphi _{\nu ^{\prime }n^{\prime }}^{\ast }(\mathbf{r})\left[ \frac{\mathbf{%
p}^{2}}{2m}-\sum_{m=1}^{N_{s}}\frac{e^{2}}{\left\vert \mathbf{r}-\mathbf{R}%
_{m}\right\vert }\right] \varphi _{\nu n}(\mathbf{r}).  \tag{3.5}
\label{3.5}
\end{equation}

Due to Eq. (2.18), this prefactor also reads%
\begin{equation}
\varepsilon (\nu ^{\prime }n^{\prime },\nu n)=\int d\mathbf{r}\text{ }%
\varphi _{\nu ^{\prime }n^{\prime }}^{\ast }(\mathbf{r})\left[ \varepsilon
_{\nu }-\sum_{m\neq n}\frac{e^{2}}{\left\vert \mathbf{r}-\mathbf{R}%
_{m}\right\vert }\right] \varphi _{\nu n}(\mathbf{r}).  \tag{3.6}
\label{3.6}
\end{equation}%
So that, for highly localized states, as the ones of physical interest, it
reduces, due to Eq. (2.20), to%
\begin{equation}
\varepsilon (\nu ^{\prime }n^{\prime },\nu n)\simeq \delta _{nn^{\prime }} 
\left[ \varepsilon _{\nu }\delta _{\nu \nu ^{\prime }}+\upsilon (\nu
^{\prime },\nu )\right] ,  \tag{3.7}  \label{3.7}
\end{equation}%
where $\upsilon (\nu ^{\prime },\nu )$ comes from the interactions with all
the other ions. Due to the translational invariance of the hamiltonian
leading to Eq. (2.18), $\upsilon (\nu ^{\prime },\nu )$ can be rewritten as%
\begin{equation}
\upsilon (\nu ^{\prime },\nu )=\sum_{m\neq n}\int d\mathbf{r}\text{ }\varphi
_{\nu ^{\prime }}^{\ast }(\mathbf{r})\sum_{m\neq n}\frac{-e^{2}}{\left\vert 
\mathbf{r}-\mathbf{(\mathbf{R}}_{m}\mathbf{-\mathbf{R}}_{n}\mathbf{)}%
\right\vert }\varphi _{\nu }(\mathbf{r})=\left\langle \nu ^{\prime
}|\sum_{R\neq 0}\frac{-e^{2}}{\left\vert \mathbf{r}-\mathbf{\mathbf{R})}%
\right\vert }|\nu \right\rangle .  \tag{3.8}  \label{3.8}
\end{equation}%
where the vectors $\mathbf{\mathbf{R}}$ in the sum correspond to all
possible distances between ions. The one-body part of the semiconductor
hamiltonian appropriate to Frenkel excitons ends by reading as%
\begin{equation}
H_{0}^{(F)}=\sum_{\nu ,n}\widetilde{\varepsilon }_{\nu }a_{\nu n}^{\dagger
}a_{\nu n}+\sum_{n,\nu ^{\prime }\neq \nu }\upsilon (\nu ^{\prime },\nu
)a_{\nu ^{\prime }n}^{\dagger }a_{\nu n}  \tag{3.9}  \label{3.9}
\end{equation}%
where $\widetilde{\varepsilon }_{\nu }=\varepsilon _{\nu }+\upsilon (\nu
,\nu )$. The second term in $H_{0}^{(F)}$ describes the fact that the
electron of a given site can change its atomic level from $\nu $ to $\nu
^{\prime }$, while staying on the same site, due to its interaction with the
ions of the other sites, as seen from the definition of $\upsilon (\nu
^{\prime },\nu )$, given by Eqs. (3.8). However, for the states $\nu =(0,1)$%
\ highly localized compared to the ion-ion distance, these $\upsilon (\nu
^{\prime },\nu )$ scatterings are extremely small.

\subsection{Electron-electron interaction}

Let us now turn to the two-body operator $V_{ee}$, defined in Eq. (2.4). The
standard second quantization procedure leads to write it on the atomic basis 
$\left\vert \nu n\right\rangle $ as 
\begin{equation}
V_{ee}=\frac{1}{2}\sum_{\left\{ \nu ,n\right\} }V\left( 
\begin{array}{cc}
\nu _{2}^{\prime }n_{2}^{\prime } & \nu _{2}n_{2} \\ 
\nu _{1}^{\prime }n_{1}^{\prime } & \nu _{1}n_{1}%
\end{array}%
\right) a_{\nu _{1}^{\prime }n_{1}^{\prime }}^{\dagger }a_{\nu _{2}^{\prime
}n_{2}^{\prime }}^{\dagger }a_{\nu _{2}n_{2}}a_{\nu _{1}n_{1}}  \tag{3.10}
\label{3.10}
\end{equation}%
where the prefactor is given by%
\begin{equation}
V\left( 
\begin{array}{cc}
\nu _{2}^{\prime }n_{2}^{\prime } & \nu _{2}n_{2} \\ 
\nu _{1}^{\prime }n_{1}^{\prime } & \nu _{1}n_{1}%
\end{array}%
\right) =\int d\mathbf{r}_{1}d\mathbf{r}_{2}\text{ }\varphi _{\nu
_{1}^{\prime }n_{1}^{\prime }}^{\ast }(\mathbf{r}_{1})\varphi _{\nu
_{2}^{\prime }n_{2}^{\prime }}^{\ast }(\mathbf{r}_{2})\frac{e^{2}}{%
\left\vert \mathbf{r}_{1}-\mathbf{r}_{2}\right\vert }\varphi _{\nu
_{2}n_{2}}(\mathbf{r}_{2})\varphi _{\nu _{1}n_{1}}(\mathbf{r}_{1}). 
\tag{3.11}  \label{3.11}
\end{equation}

For $\nu $ and $\nu ^{\prime }$ equal to $(0,1)$, \ this prefactor is
nonzero for $n_{1}^{\prime }=n_{1}$ and $n_{2}^{\prime }=n_{2}$ only, due to
Eq. (2.20). If we then use the translational invariance of atomic wave
functions, namely, Eq. (2.18), it is easy to see that $V_{ee}$ can be
written as 
\begin{equation}
V_{ee}=\frac{1}{2}\sum_{\substack{ n_{1}\nu _{1}^{\prime }\nu _{1}  \\ %
n_{2}\nu _{2}^{\prime }\nu _{2}}}V_{R_{n_{1}}-R_{n_{2}}}\left( 
\begin{array}{cc}
\nu _{2}^{\prime } & \nu _{2} \\ 
\nu _{1}^{\prime } & \nu _{1}%
\end{array}%
\right) a_{\nu _{1}^{\prime }n_{1}}^{\dagger }a_{\nu _{2}^{\prime
}n_{2}}^{\dagger }a_{\nu _{2}n_{2}}a_{\nu _{1}n_{1}}  \tag{3.12}
\label{3.12}
\end{equation}%
where the electron-electron scattering depends on the distance $R$ between
ions through%
\begin{equation}
V_{R}\left( 
\begin{array}{cc}
\nu _{2}^{\prime } & \nu _{2} \\ 
\nu _{1}^{\prime } & \nu _{1}%
\end{array}%
\right) =V_{-R}\left( 
\begin{array}{cc}
\nu _{2}^{\prime } & \nu _{2} \\ 
\nu _{1}^{\prime } & \nu _{1}%
\end{array}%
\right) =\int d\mathbf{r}_{1}d\mathbf{r}_{2}\varphi _{\nu _{1}^{\prime
}}^{\ast }(\mathbf{r}_{1})\varphi _{\nu _{2}^{\prime }}^{\ast }(\mathbf{r}%
_{2})\frac{e^{2}}{\left\vert \mathbf{r}_{1}-\mathbf{r}_{2}+\mathbf{R}%
\right\vert }\varphi _{\nu _{2}}(\mathbf{r}_{2})\varphi _{\nu _{1}}(\mathbf{r%
}_{1}).  \tag{3.13}  \label{3.13}
\end{equation}

\subsection{Restricted hamiltonian for Frenkel excitons}

The atomic states ($\nu =0,n$) and ($\nu =1,n$) are the equivalents of the
valence and the conduction band states for Wannier exciton. By noting that
the energy of the atomic ground state $\varepsilon _{\nu =0}$ is very
different from the one of the first excited level $\varepsilon _{\nu =1}$,
we are led to think that the physically relevant part of the hamiltonian
corresponds to processes in which the number of electrons in the $\nu =0$
level and the number of electrons in the $\nu =1$ level are separately
conserved. This is equivalent to processes which keep the number of
conduction electrons and the number of valence electrons fixed, for the case
of Wannier excitons. This leads us to drop terms with $\nu \neq \nu ^{\prime
}$ in $H_{0}^{(F)}$. The latter then reduces to the first sum of Eq. (3.9),
namely%
\begin{equation}
H_{0}^{(F)}\simeq H_{0}=\widetilde{\varepsilon }_{0}\sum_{n}a_{0n}^{\dagger
}a_{0n}+\widetilde{\varepsilon }_{1}\sum_{n}a_{1n}^{\dagger }a_{1n}. 
\tag{3.14}  \label{3.14}
\end{equation}

If we now turn to the Coulomb interaction $V_{ee}$, given in Eq. (3.12), and
we also keep terms which conserve the number of $\nu =0$ and $\nu =1$
electrons separately, we see that they are of two kinds. (i) $V_{ee}$
contains intraatomic processes, in which the electron of a given site stays
on the same level. These processes correspond to terms like $a_{\nu
n_{1}}^{\dagger }a_{\nu n_{2}}^{\dagger }a_{\nu n_{2}}a_{\nu n_{1}}$with $%
\nu $ equal 0 or 1. $V_{ee}$ also contains terms like\ $a_{0n_{1}}^{\dagger
}a_{1n_{2}}^{\dagger }a_{1n_{2}}a_{0n_{1}}$ with 0 and 1 possibly exchanged
- which makes such a term appearing with a factor of 2. (ii) In addition, $%
V_{ee}$ contains interatomic processes, in which the electron of one site
jumps from $\nu =0$ to $\nu =1$, while the electron of another site goes
from $\nu =1$ to $\nu =0$, namely, terms like $a_{0n_{1}}^{\dagger
}a_{1n_{2}}^{\dagger }a_{0n_{2}}a_{1n_{1}}$, with 0 and 1 possibly exchanged
- which makes this term also appearing with a factor of 2.

So that $V_{ee}$ for Frenkel exciton ultimately reduces to $V_{ee}\simeq
V_{ee}^{(1)}+V_{ee}^{(2)}+V_{ee}^{(3)}+V_{ee}^{(4)}$ with\bigskip 
\begin{equation*}
V_{ee}^{(1)}=\frac{1}{2}\sum_{n_{1}\neq n_{2}}V_{R_{n_{1}}-R_{n_{2}}}\left( 
\begin{array}{cc}
0 & 0 \\ 
0 & 0%
\end{array}%
\right) a_{0n_{1}}^{\dagger }a_{0n_{2}}^{\dagger }a_{0n_{2}}a_{0n_{1}},
\end{equation*}%
\begin{equation*}
V_{ee}^{(2)}=\frac{1}{2}\sum_{n_{1}\neq n_{2}}V_{R_{n_{1}}-R_{n_{2}}}\left( 
\begin{array}{cc}
1 & 1 \\ 
1 & 1%
\end{array}%
\right) a_{1n_{1}}^{\dagger }a_{1n_{2}}^{\dagger }a_{1n_{2}}a_{1n_{1}},
\end{equation*}%
\begin{equation*}
V_{ee}^{(3)}=\sum_{n_{1}n_{2}}V_{R_{n_{1}}-R_{n_{2}}}\left( 
\begin{array}{cc}
1 & 1 \\ 
0 & 0%
\end{array}%
\right) a_{0n_{1}}^{\dagger }a_{1n_{2}}^{\dagger }a_{1n_{2}}a_{0n_{1}},
\end{equation*}%
\begin{equation}
V_{ee}^{(4)}=\sum_{n_{1}n_{2}}V_{R_{n_{1}}-R_{n_{2}}}\left( 
\begin{array}{cc}
0 & 1 \\ 
1 & 0%
\end{array}%
\right) a_{1n_{1}}^{\dagger }a_{0n_{2}}^{\dagger }a_{1n_{2}}a_{0n_{1}}. 
\tag{3.15}  \label{3.15}
\end{equation}

Note that the last two terms differ from zero for $n_{1}=n_{2}$, while the
first two terms are equal to zero for $n_{1}=n_{2}$; this is why we have
excluded $n_{1}=n_{2}$ from the first two sums.

These four terms are shown in Fig. 1.

\section{Electron-hole hamiltonian for Frenkel excitons}

\subsection{Electron and hole creation operators}

As for Wannier excitons, it is appropriate to introduce the concept of hole.
This will allow us to start with a $\left\vert 0\right\rangle $ state, in
which the electrons of all sites are in the atomic ground state, $\nu =0$,
and to speak in terms of excitations with respect to this ground state,
i.e., in terms of the small number of sites in which the electron is no more
in the ground state, this number being one for one exciton, two for two
excitons and so on... Such an elementary electron-hole excitation is shown
in Fig. 2. Due to the electron-hole attraction and the cost in electrostatic
energy induced by the electron and hole separation, we expect the lowest
energy excited states to correspond to $n_{1}=n_{2}$. As a direct
consequence, the Frenkel excitons are going to be made from electron-hole
pairs on the same site. Let us now recover this obvious result.

In the absence of spin degrees of freedom, the electron and hole creation
operators are simply linked to the $\nu =0$ and $\nu =1$ atomic state level
creation operators through%
\begin{equation*}
a_{1n}^{\dagger }=a_{n}^{\dagger }
\end{equation*}
\begin{equation}
a_{0n}=b_{n}^{\dagger }.  \tag{4.1}  \label{4.1}
\end{equation}

By using the anticommutation relations for electrons in atomic states, given
by Eq. (3.3), it is straightforward to show that $\left[ a_{n^{\prime
}},a_{n}\right] _{+}=0$, while 
\begin{equation}
\left[ a_{n^{\prime }},a_{n}^{\dagger }\right] _{+}=\left[ a_{1n^{\prime
}},a_{1n}^{\dagger }\right] _{+}=\left\langle 1n^{\prime }|1n\right\rangle
\simeq \delta _{nn^{\prime }}  \tag{4.2}  \label{4.2}
\end{equation}%
for highly localized atomic states compared to the interatomic distance,
which makes Eq. (2.21) valid. In the same way, $\left[ b_{n^{\prime }},b_{n}%
\right] _{+}=0$, while%
\begin{equation}
\left[ b_{n^{\prime }},b_{n}^{\dagger }\right] _{+}=\left[ a_{0n^{\prime
}}^{\dagger },a_{0n}\right] _{+}=\left\langle 0n^{\prime }|0n\right\rangle
\simeq \delta _{n^{\prime }n}.  \tag{4.3}  \label{4.3}
\end{equation}

If we now turn to the anticommutator between electron and hole operators, we
find that%
\begin{equation*}
\left[ a_{n^{\prime }},b_{n}^{\dagger }\right] _{+}=\left[ a_{1n^{\prime
}},a_{0n}\right] _{+}=0,
\end{equation*}%
\begin{equation}
\left[ a_{n^{\prime }},b_{n}\right] _{+}=\left[ a_{1n^{\prime
}},a_{0n}^{\dagger }\right] _{+}=\left\langle 1n^{\prime }|0n\right\rangle
\simeq 0,  \tag{4.4}  \label{4.4}
\end{equation}%
for highly localized atomic states, this last anticommutator being exactly
equal to zero for $n=n^{\prime }$, due to Eq. (2.19a).

\subsection{One-body operator $H_{0}$ in terms of electrons and holes}

In order to rewrite the part $H_{0}$ of the Frenkel exciton hamiltonian
given in Eq. (3.14) in terms of electron and hole operators, we first note
that $a_{0n}^{\dagger }a_{0n}=1-a_{0n}a_{0n}^{\dagger }=1-b_{n}^{\dagger
}b_{n}$; so that%
\begin{equation}
H_{0}=N_{s}\widetilde{\varepsilon }_{0}+\sum_{n}(\widetilde{\varepsilon }%
_{1}a_{n}^{\dagger }a_{n}-\widetilde{\varepsilon }_{0}b_{n}^{\dagger }b_{n}).
\tag{4.5}  \label{4.5}
\end{equation}%
Due to other contributions in $a_{n}^{\dagger }a_{n}$ and $b_{n}^{\dagger
}b_{n}$\ coming from the electron-electron interaction, the electron energy
and the hole energy are going to differ from $\widetilde{\varepsilon }_{1}$
and $(-\widetilde{\varepsilon }_{0})$, as now shown.

\subsection{Electron-electron interaction in terms of electrons and holes}

We now turn to the electron-electron interactions given in Eq. (3.15). Since 
$a_{1n}=a_{n}$, the second term of this equation, shown in Fig. 3a, readily
gives the Coulomb repulsion between two electrons as 
\begin{equation}
V_{ee}^{(2)}=\widetilde{V}_{ee}=\frac{1}{2}\sum_{n_{1}\neq
n_{2}}V_{R_{n_{1}}-R_{n_{2}}}\left( 
\begin{array}{cc}
1 & 1 \\ 
1 & 1%
\end{array}%
\right) a_{n_{1}}^{\dagger }a_{n_{2}}^{\dagger }a_{n_{2}}a_{n_{1}}. 
\tag{4.6}  \label{4.6}
\end{equation}

In order to rewrite the third term of Eq. (3.15), we first note that $%
a_{0n_{1}}^{\dagger }a_{1n_{2}}^{\dagger
}a_{1n_{2}}a_{0n1}=(-a_{1n_{2}}^{\dagger }a_{0n_{1}}^{\dagger
})(-a_{0n_{1}}a_{1n_{2}})$, while $a_{0n_{1}}^{\dagger
}a_{0n_{1}}=1-a_{0n_{1}}a_{0n_{1}}^{\dagger }$; so that this third term
gives two contributions,%
\begin{equation}
V_{ee}^{(3)}=V_{eh}^{(dir)}+\sum_{n_{2}}a_{n_{2}}^{\dagger
}a_{n_{2}}\sum_{n_{1}}V_{R_{n_{1}}-R_{n_{2}}}\left( 
\begin{array}{cc}
1 & 1 \\ 
0 & 0%
\end{array}%
\right) .  \tag{4.7}  \label{4.7}
\end{equation}%
The second term of $V_{ee}^{(3)}$, which comes from the Coulomb interaction
between one electron in an atomic excited level and all the atomic ground
states, is going to dress the bare electron energy $\widetilde{\varepsilon }%
_{1}$, appearing in Eq. (4.5). The first term of $V_{ee}^{(3)}$, given by%
\begin{equation}
V_{eh}^{(dir)}=-\sum_{n_{1}n_{2}}V_{R_{n_{1}}-R_{n_{2}}}\left( 
\begin{array}{cc}
1 & 1 \\ 
0 & 0%
\end{array}%
\right) b_{n_{1}}^{\dagger }a_{n_{2}}^{\dagger }a_{n_{2}}b_{n_{1}}, 
\tag{4.8}  \label{4.8}
\end{equation}%
describes a direct electron-hole attraction, the electron and the hole
staying in their sites (see Fig (3c)).

In the same way, the fourth term of Eq. (3.15) gives two contributions since 
$a_{1n_{1}}^{\dagger }a_{0n_{2}}^{\dagger
}a_{1n_{2}}a_{0n_{1}}=a_{1n_{1}}^{\dagger }a_{0n_{2}}^{\dagger
}(-a_{0n_{1}}a_{1n_{2}})$ while\ $-a_{0n_{2}}^{\dagger }a_{0n_{1}}=-\delta
_{n_{1}n_{2}}+a_{0n_{1}}a_{0n_{2}}^{\dagger }$; so that it reads 
\begin{equation}
V_{ee}^{(4)}=V_{eh}^{(exch)}-V_{R=0}\left( 
\begin{array}{cc}
0 & 1 \\ 
1 & 0%
\end{array}%
\right) \sum_{n}a_{n}^{\dagger }a_{n}.  \tag{4.9}  \label{4.9}
\end{equation}%
The second term of $V_{ee}^{(4)}$ is also going to dress the electron energy 
$\widetilde{\varepsilon }_{1}$, while the first term, given by 
\begin{equation}
V_{eh}^{(exch)}=\sum_{n_{1}n_{2}}V_{R_{n_{1}}-R_{n_{2}}}\left( 
\begin{array}{cc}
0 & 1 \\ 
1 & 0%
\end{array}%
\right) a_{n_{1}}^{\dagger }b_{n_{1}}^{\dagger }b_{n_{2}}a_{n_{2}} 
\tag{4.10}  \label{4.10}
\end{equation}%
and shown in Fig. (3d), corresponds to the destruction of one electron-hole
pair on the site $n_{2}$ and to its recreation on the site $n_{1}$. Let us
stress that, while the direct electron-hole Coulomb interaction $%
V_{eh}^{(dir)}$ in Eq. (4.8) corresponds to an attraction, the exchange
electron-hole Coulomb interaction $V_{eh}^{(exch)}$ is repulsive.

We now turn to the first term of Eq. (3.15) between two ground state
electrons $\nu =0$. We first note that, due to Eqs. (3.3-4),%
\begin{equation}
a_{0n_{1}}^{\dagger }a_{0n_{2}}^{\dagger
}a_{0n_{2}}a_{0n_{1}}=1-a_{0n_{1}}a_{0n_{1}}^{\dagger
}-a_{0n_{2}}a_{0n_{2}}^{\dagger }+a_{0n_{2}}a_{0n_{1}}a_{0n_{1}}^{\dagger
}a_{0n_{2}}^{\dagger }.  \tag{4.11}  \label{4.11}
\end{equation}%
So that, by grouping the two terms with a minus sign, $V_{ee}^{(1)}$
generates three contributions%
\begin{equation}
V_{ee}^{(1)}=\frac{1}{2}\sum_{n_{1}\neq n_{2}}V_{R_{n_{1}}-R_{n_{2}}}\left( 
\begin{array}{cc}
0 & 0 \\ 
0 & 0%
\end{array}%
\right) -\sum_{n}b_{n}^{\dagger }b_{n}\sum_{n^{\prime }\neq
n}V_{R_{n}-R_{n^{\prime }}}\left( 
\begin{array}{cc}
0 & 0 \\ 
0 & 0%
\end{array}%
\right) +\widetilde{V}_{hh}.  \tag{4.12}  \label{4.12}
\end{equation}

The last term of $V_{ee}^{(1)}$, shown in Fig. (3b), corresponds to an
hole-hole repulsion. It precisely reads

\begin{equation}
\widetilde{V}_{hh}=\frac{1}{2}\sum_{n_{1}\neq
n_{2}}V_{R_{n_{1}}-R_{n_{2}}}\left( 
\begin{array}{cc}
0 & 0 \\ 
0 & 0%
\end{array}%
\right) b_{n_{1}}^{\dagger }b_{n_{2}}^{\dagger }b_{n_{2}}b_{n_{1}}. 
\tag{4.13}  \label{4.13}
\end{equation}%
The first term of $V_{ee}^{(1)}$ is a bare constant which describes all
Coulomb interactions between ground state atomic levels. It produces a band
gap renormalization. The second term of $V_{ee}^{(1)}$ comes from the
interaction between one particular ground state electron, in the site $n$,
and the other ground state electrons. This term has to appear when the site $%
n$ is empty, i.e., occupied by a hole, in order to compensate for the
electron-electron interaction already included in the constant term of $%
V_{ee}^{(1)}$. This second term is going to dress the atomic ground state
energy $\widetilde{\varepsilon }_{0}$, when speaking in terms of holes, as
fully reasonable, since all interactions between atomic ground state
electrons are by construction forgotten when we turn to electrons and holes.
These interactions actually appear through the renormalization of the atomic
ground state and excited state energies, the electron and the hole being
more subtle objects that just one electron in the atomic excited state and
one electron absence in the atomic ground state.

\subsection{Electron-hole hamiltonian}

If we now collect all these terms, we end by writing the part of the
semiconductor hamiltonian, appropriate to Frenkel excitons $%
H_{0}+V_{ee}^{(1)}+V_{ee}^{(2)}+V_{ee}^{(3)}+V_{ee}^{(4)}+V_{ion-ion}$ as

\begin{equation}
H^{(F)}=\Delta +H_{eh}+V_{intra}+V_{inter}.  \tag{4.14}  \label{4.14}
\end{equation}

(i) $\Delta $ is a constant which contains contributions from the atomic
level ground states only. It precisely reads%
\begin{equation}
\Delta =N_{s}\varepsilon _{0}+N_{s}\upsilon (0,0)+\frac{1}{2}\sum_{n_{1}\neq
n_{2}}V_{R_{n_{1}}-R_{n_{2}}}\left( 
\begin{array}{cc}
0 & 0 \\ 
0 & 0%
\end{array}%
\right) +\frac{1}{2}\sum_{n_{1}\neq n_{2}}\frac{e^{2}}{\left\vert \mathbf{R}%
_{n_{1}}-\mathbf{R}_{n_{2}}\right\vert }=N_{s}(\varepsilon _{0}+\varepsilon
_{0}^{(coul)})  \tag{4.15}  \label{4.15}
\end{equation}%
where $N_{s}$ is the number of ion sites. By using Eqs. (3.8) and (3.13),
the Coulomb contribution to this band gap renormalization, given by the
bracket of the above expression, can be rewritten as%
\begin{equation}
\varepsilon _{0}^{(coul)}=\sum_{R\neq 0}\int d\mathbf{r}d\mathbf{r}^{\prime
}\left\vert \varphi _{0}(\mathbf{r})\right\vert ^{2}\left\vert \varphi _{0}(%
\mathbf{r}^{\prime })\right\vert ^{2}\left[ \frac{-e^{2}}{\left\vert \mathbf{%
r}-\mathbf{R}\right\vert }+\frac{1}{2}\frac{e^{2}}{\left\vert \mathbf{r}-%
\mathbf{r}^{\prime }-\mathbf{R}\right\vert }+\frac{1}{2}\frac{e^{2}}{R}%
\right]   \tag{4.16}  \label{4.16}
\end{equation}%
where the $R$'s are the possible distances between two ions. Note that the
last term, $e^{2}/R$, which comes from the ion-ion interaction, and which
makes the system at hand neutral, allows for the convergence of $\varepsilon
_{0}^{(coul)}$ in the large sample limit.

(ii) The second term of Eq. (4.14) is a one-body operator which can be
written as%
\begin{equation}
H_{eh}=\varepsilon _{e}\sum_{n}a_{n}^{\dagger }a_{n}+\varepsilon
_{h}\sum_{n}b_{n}^{\dagger }b_{n}.  \tag{4.17}  \label{4.17}
\end{equation}%
It describes the electron and hole kinetic energies. These energies, given by%
\begin{equation}
\varepsilon _{e}=\varepsilon _{1}+\upsilon (1,1)+\sum_{R}V_{R}\left( 
\begin{array}{cc}
1 & 1 \\ 
0 & 0%
\end{array}%
\right) -V_{R=0}\left( 
\begin{array}{cc}
0 & 1 \\ 
1 & 0%
\end{array}%
\right) ,  \tag{4.18}  \label{4.18}
\end{equation}

\begin{equation}
-\varepsilon _{h}=\varepsilon _{0}+\upsilon (0,0)+\sum_{R\neq 0}V_{R}\left( 
\begin{array}{cc}
0 & 0 \\ 
0 & 0%
\end{array}%
\right)  \tag{4.19}  \label{4.19}
\end{equation}%
differ from the atomic bare ground and excited state energies $-\varepsilon
_{0}$ and $\varepsilon _{1}$ due to Coulomb interactions with all the atomic
ground states. These contributions have to appear, when we speak in terms of
holes since all Coulomb interactions among these atomic ground state levels
are then forgotten, by construction.

(iii) The third term $V_{intra}$ of Eq. (4.14) corresponds to $%
V_{eh}^{(dir)}+V_{eh}^{(exch)}$ taken for $n_{1}=n_{2}$. It precisely reads%
\begin{equation}
V_{intra}=-\delta \sum_{n}a_{n}^{\dagger }b_{n}^{\dagger }b_{n}a_{n}, 
\tag{4.20}  \label{4.20}
\end{equation}%
where $(-\delta )$ is given by%
\begin{equation}
-\delta =-V_{R=0}\left( 
\begin{array}{cc}
1 & 1 \\ 
0 & 0%
\end{array}%
\right) +V_{R=0}\left( 
\begin{array}{cc}
0 & 1 \\ 
1 & 0%
\end{array}%
\right) .  \tag{4.21}  \label{4.21}
\end{equation}

By using Coulomb couplings, given in Eq. (3.13), we see that this quantity
also reads%
\begin{equation}
\delta =\int d\mathbf{r}_{1}d\mathbf{r}_{2}\left[ \varphi _{_{1}}^{\ast }(%
\mathbf{r}_{2})\varphi _{0}^{\ast }(\mathbf{r}_{1})-\varphi _{_{0}}^{\ast }(%
\mathbf{r}_{2})\varphi _{1}^{\ast }(\mathbf{r}_{1})\right] \frac{e^{2}}{%
\left\vert \mathbf{r}_{1}-\mathbf{r}_{2}\right\vert }\varphi _{_{1}}^{\ast }(%
\mathbf{r}_{2})\varphi _{0}^{\ast }(\mathbf{r}_{1}).  \tag{4.22}
\label{4.22}
\end{equation}%
This shows that $\delta $\ is a positive constant, since $\left\langle \nu
|\nu \right\rangle =1$, while $\left\langle 0|1\right\rangle =0$. This
energy $\delta $ physically corresponds to the energy decrease when the site 
$n$ is occupied by an electron-hole pair, i.e., when the site $n$ is
neutral. This is going to make the potential $V_{intra}$ responsible for the
fact that excitons are made from linear combinations of electrons and holes
located on the same site.

(iv) The last term of Eq. (4.14), $V_{inter}$, is made of all Coulomb
interactions \textit{between} sites. It contains the electron-electron and
hole-hole contributions $\widetilde{V}_{ee}$ and $\widetilde{V}_{hh}$, given
by Eqs. (4.6) and (4.13), which are interactions between sites by
construction, since a given site cannot accommodate two electrons or two
holes due to the Pauli exclusion principle. It also contains the part of the
direct electron-hole potential $V_{eh}^{(dir)}$, taken for $n_{1}\neq n_{2}$%
. Using Eq. (4.8), this direct electron-hole exchange interaction between
sites precisely reads%
\begin{equation}
\widetilde{V}_{eh}^{(dir)}=-\sum_{n_{1}\neq
n_{2}}V_{R_{n_{1}}-R_{n_{_{2}}}}\left( 
\begin{array}{cc}
1 & 1 \\ 
0 & 0%
\end{array}%
\right) b_{n_{1}}^{\dagger }a_{n_{2}}^{\dagger }a_{n_{2}}b_{n_{1}}. 
\tag{4.23}  \label{4.23}
\end{equation}%
It finally contains the part of the electron-hole exchange potential $%
V_{eh}^{(exch)}$ given in Eq. (4.10), taken for $n_{1}\neq n_{2}$. This part
has a very special role since it allows the excitation transfer from one
site to the other. Let us isolate this transfer term from the other Coulomb
terms and call it as $V_{trans}$%
\begin{equation}
V_{trans}=\sum_{n_{1}\neq n_{2}}V_{R_{n_{1}}-R_{n_{2}}}\left( 
\begin{array}{cc}
0 & 1 \\ 
1 & 0%
\end{array}%
\right) a_{n_{2}}^{\dagger }b_{n_{2}}^{\dagger }b_{n_{1}}a_{n_{1}}. 
\tag{4.24}  \label{4.24}
\end{equation}%
All this leads us to write $V_{inter}$ in Eq. (4.14) as%
\begin{equation*}
V_{inter}=V_{trans}+V_{coul},
\end{equation*}%
\begin{equation}
V_{coul}=\widetilde{V}_{ee}+\widetilde{V}_{hh}+\widetilde{V}_{eh}^{(dir)}, 
\tag{4.25}  \label{4.25}
\end{equation}

The four contributions of this $V_{inter}$ operator are shown in Fig. (4).

\subsection{Discussion}

The expression of the semiconductor hamiltonian appropriate to Frenkel
exciton $H^{(F)}$ in terms of electrons and holes, given in Eq. (4.14),
allows an easy comparison between highly localized states leading to Frenkel
excitons, and extended states, leading to Wannier excitons. We first see
that the electron and the hole energies for Frenkel excitons appearing in $%
H_{eh}$ are constant, while they depend on $\mathbf{k}$ for Wannier exciton:
the electrons and holes for Wannier excitons, which belong to the conduction
and valence bands, are delocalized over the whole sample, so that their
energies must depend on momentum. In addition, the "electon-hole exchange",
i.e. the possibility for one electron-hole pair to recombine while another
pair is created, plays essentially no role for Wannier excitons: it is just
responsible for a small splitting between Wannier excitons, when the spin
degrees of freedom are included. On the opposite, this "electron-hole
exchange" is crucial in the case of highly localized states, as it is the
only process allowing an excitation transfer between sites: this makes the
operator $V_{trans}$ entirely responsible for the Frenkel exciton formation,
as we now show.

\section{Lowest excited states in the absence of interaction between sites}

Let us first forget the interactions between the sites. The hamiltonian $%
H^{(F)}$, given in Eq. (4.14), then reduces to 
\begin{equation}
H_{pair}=H_{eh}+V_{intra},  \tag{5.1}  \label{5.1}
\end{equation}%
if we drop the irrelevant band gap renormalization $\Delta $.

\subsection{Ground state and lowest excited states}

The ground state of $H_{pair}$ has zero electron-hole pair. Let us call it $%
\left\vert 0\right\rangle $ and take its energy as 0.

If we now consider the one electron-hole pair state $a_{n}^{\dagger
}b_{n^{\prime }}^{\dagger }\left\vert 0\right\rangle $ with an electron on
site $n$ and a hole on site $n^{\prime }$, we see that its energy is $%
\varepsilon _{e}+\varepsilon _{h}$ for $n\neq n^{\prime }$ while it is $%
\varepsilon _{e}+\varepsilon _{h}-\delta $ for $n=n^{\prime }$. Since $%
\delta $ is positive, the lowest excited states of $H_{pair}$ thus have one
electron-hole pair on the same site. They reads%
\begin{equation*}
(H_{pair}-E_{pair})\left\vert R_{n}\right\rangle =0,
\end{equation*}%
\begin{equation}
\left\vert R_{n}\right\rangle =a_{n}^{\dagger }b_{n}^{\dagger }\left\vert
0\right\rangle =B_{n}^{\dagger }\left\vert 0\right\rangle  \tag{5.2}
\label{5.2}
\end{equation}%
with $E_{pair}=\varepsilon _{e}+\varepsilon _{h}-\delta $. These states form
a $N_{s}$-degenerate subspace, since $n$ can run from 1 to $N_{s}$.

\subsection{Commutation rules}

Using the anticommutation rules for electrons and holes given in Eqs.
(4.2-4), it is easy to show that the electron-hole pair operators $%
B_{n}^{\dagger }$ behave as bosons with respect to the destruction operators
since their commutator reads%
\begin{equation}
\left[ B_{n},B_{n^{\prime }}\right] _{-}=0,  \tag{5.3}  \label{5.3}
\end{equation}%
while they are composite bosons only since the other commutator is such that%
\begin{equation}
\left[ B_{n^{\prime }},B_{n}^{\dagger }\right] _{-}=\delta _{nn^{\prime
}}-D_{nn^{\prime }},  \tag{5.4}  \label{5.4}
\end{equation}%
the deviation-from-boson operator for electron-hole pairs being equal to%
\begin{equation}
D_{nn^{\prime }}=\delta _{nn^{\prime }}(a_{n}^{\dagger }a_{n}+b_{n}^{\dagger
}b_{n}).  \tag{5.5}  \label{5.5}
\end{equation}

As standard for deviation-from-boson operator, $D_{nn^{\prime }}$ gives 0
when acting on the electron-hole pair vacuum.

\section{Frenkel excitons}

If we now keep the coupling between sites $V_{inter}=V_{trans}+V_{coul}$,
defined in Eq. (4.25), in the Frenkel exciton hamiltonian $H^{(F)}$, we
induce non-diagonal contributions between different sites. They are going to
split the degenerate subspace $\left\vert R_{n}\right\rangle $. The Frenkel
excitons result from the diagonalization of the hamiltonian $H^{(F)}$ in
this $\left\vert R_{n}\right\rangle $ degenerate subspace.

\subsection{Derivation of the Frenkel excitons}

Since the states $\left\vert R_{n}\right\rangle $ have one electron-hole
pair only, the electron-electron and hole-hole parts $\widetilde{V}_{ee}$
and $\widetilde{V}_{hh}$ of $V_{coul}$, defined in Eq. (4.25), give zero
when acting on $\left\vert R_{n}\right\rangle $. The direct electron-hole
interaction $\widetilde{V}_{eh}^{(dir)}$, given in Eq. (4.23), also gives
zero since $b_{n_{2}}a_{n_{1}}\left\vert R_{n}\right\rangle =0$ for $%
n_{1}\neq n_{2}$. Consequently $V_{coul}\left\vert R_{n}\right\rangle =0$;
so that the only part of $V_{inter}$, which plays a role in the
diagonalization of $H^{(F)}$ in the $\left\vert R_{n}\right\rangle $
subspace is the electron-hole exchange term $V_{trans}$. Since $%
b_{n_{2}}a_{n_{2}}a_{n}^{\dagger }b_{n}^{\dagger }\left\vert 0\right\rangle
=\delta _{n_{2}n}\left\vert 0\right\rangle $, we readily find%
\begin{equation}
V_{trans}\left\vert R_{n}\right\rangle =\sum_{n_{1}\neq
n}V_{R_{n_{1}}-R_{n}}\left( 
\begin{array}{cc}
0 & 1 \\ 
1 & 0%
\end{array}%
\right) \left\vert R_{n_{1}}\right\rangle .  \tag{6.1}  \label{6.1}
\end{equation}

This shows that, if we drop the irrelevant constant $\Delta $, the Frenkel
exciton hamiltonian $H^{(F)}$ acting on $\left\vert R_{n}\right\rangle $
reduces to%
\begin{equation*}
H_{X}^{(0)}=H_{pair}+V_{trans},
\end{equation*}
\begin{equation}
H_{X}^{(0)}\left\vert R_{n}\right\rangle =E_{pair}\left\vert
R_{n}\right\rangle +\sum_{n_{1}\neq n}V_{R_{n_{1}}-R_{n}}\left( 
\begin{array}{cc}
0 & 1 \\ 
1 & 0%
\end{array}%
\right) \left\vert R_{n_{1}}\right\rangle .  \tag{6.2}  \label{6.2}
\end{equation}

Let us now show that the following linear combinations of $\left\vert
R_{n}\right\rangle $ 
\begin{equation}
\left\vert X_{Q}\right\rangle =\frac{1}{\sqrt{N_{s}}}\sum_{n=1}^{N_{s}}e^{i%
\mathbf{Q.R}_{n}}\left\vert R_{n}\right\rangle ,  \tag{6.3}  \label{6.3}
\end{equation}%
known as Frenkel excitons, are the exact eigenstates of the hamiltonian $%
H_{x}^{(0)}$. To prove it, we first consider $V_{trans}$ acting on $%
\left\vert X_{Q}\right\rangle $ 
\begin{equation}
V_{trans}\left\vert X_{Q}\right\rangle =\frac{1}{\sqrt{N_{s}}}\sum_{n}e^{i%
\mathbf{Q.R}_{n}}\sum_{n^{\prime }\neq n}V_{R_{n}-R_{n^{\prime }}}\left( 
\begin{array}{cc}
0 & 1 \\ 
1 & 0%
\end{array}%
\right) \left\vert R_{n^{\prime }}\right\rangle ,  \tag{6.4}  \label{6.4}
\end{equation}%
which also reads 
\begin{equation}
V_{trans}\left\vert X_{Q}\right\rangle =\frac{1}{\sqrt{N_{s}}}%
\sum_{n^{\prime }}\left\vert R_{n^{\prime }}\right\rangle \sum_{n\neq
n^{\prime }}e^{i\mathbf{Q.\mathbf{R}}_{n}}V_{R_{n}-R_{n^{\prime }}}\left( 
\begin{array}{cc}
0 & 1 \\ 
1 & 0%
\end{array}%
\right) .  \tag{6.5}  \label{6.5}
\end{equation}

In order to calculate the last sum, we rewrite it as%
\begin{equation}
\sum_{n\neq n^{\prime }}e^{i\mathbf{Q.\mathbf{R}}_{n}}V_{R_{n}-R_{n^{\prime
}}}\left( 
\begin{array}{cc}
0 & 1 \\ 
1 & 0%
\end{array}%
\right) =e^{i\mathbf{Q.\mathbf{R}}_{n^{\prime }}}\sum_{n\neq n^{\prime }}e^{i%
\mathbf{Q.(\mathbf{\mathbf{R}}}_{n}-\mathbf{\mathbf{\mathbf{R}}}_{n^{\prime
}}\mathbf{)}}V_{R_{n}-R_{n^{\prime }}}\left( 
\begin{array}{cc}
0 & 1 \\ 
1 & 0%
\end{array}%
\right) .  \tag{6.6}  \label{6.6}
\end{equation}

Due to the invariance of the system, the sum in RHS cannot depend on $%
n^{\prime }$; so that the above equation leads to%
\begin{equation}
\sum_{n\neq n^{\prime }}e^{i\mathbf{Q.\mathbf{R}}_{n}}V_{R_{n^{\prime
}}-R_{n}}\left( 
\begin{array}{cc}
0 & 1 \\ 
1 & 0%
\end{array}%
\right) =e^{i\mathbf{Q.\mathbf{R}}_{n^{\prime }}}\sum_{R\neq 0}e^{-i\mathbf{%
Q.\mathbf{\mathbf{R}}}}V_{R}\left( 
\begin{array}{cc}
0 & 1 \\ 
1 & 0%
\end{array}%
\right) ,  \tag{6.7}  \label{6.7}
\end{equation}%
where the sum is taken over all possible distances $R$ between ions. When
inserted into Eq. (6.5), this readily leads to%
\begin{equation}
V_{trans}\left\vert X_{Q}\right\rangle =\left\vert X_{Q}\right\rangle
\sum_{R\neq 0}e^{-i\mathbf{Q.\mathbf{\mathbf{R}}}}V_{R}\left( 
\begin{array}{cc}
0 & 1 \\ 
1 & 0%
\end{array}%
\right) .  \tag{6.8}  \label{6.8}
\end{equation}

So that we end with 
\begin{equation}
H_{X}^{(0)}\left\vert X_{Q}\right\rangle =E_{Q}\left\vert X_{Q}\right\rangle
,  \tag{6.9}  \label{6.9}
\end{equation}%
where the eigenenergy is given by%
\begin{equation}
E_{Q}=E_{pair}+\sum_{R\neq 0}e^{-i\mathbf{Q.R}}V_{R}\left( 
\begin{array}{cc}
0 & 1 \\ 
1 & 0%
\end{array}%
\right) .  \tag{6.10}  \label{6.10}
\end{equation}%
The above equation shows that the splitting of the $N_{s}$ degenerate states 
$\left\vert R_{n}\right\rangle $ into $N_{s}$ states $\left\vert
X_{Q}\right\rangle $ is only due to the electron-hole exchange $V_{trans}$
between different sites: $V_{trans}$ is the only part of the hamiltonian
allowing for an excitation transfer from site to site, as necessary for the
delocalization of the Frenkel excitons.

\subsection{Frenkel exciton creation operator}

Eq. (6.3) leads to write the Frenkel exciton creation operator $%
B_{Q}^{\dagger }$ defined as $\left\vert X_{Q}\right\rangle =B_{Q}^{\dagger
}\left\vert 0\right\rangle $ in terms of the creation operators $%
B_{n}^{\dagger }=a_{n}^{\dagger }b_{n}^{\dagger }$ for one electron-hole
pair on site $n$ as 
\begin{equation}
B_{Q}^{\dagger }=\frac{1}{\sqrt{N_{s}}}\sum_{n=1}^{N_{s}}e^{i\mathbf{Q.R}%
_{n}}B_{n}^{\dagger }.  \tag{6.11}  \label{6.11}
\end{equation}%
In the same way as free electron-hole pairs can be written in terms of
Wannier excitons, it is possible to write electron-hole pair localized on
site $n$ in terms of Frenkel excitons. This is barely done by noting that%
\begin{equation}
\frac{1}{\sqrt{N_{s}}}\sum_{\mathbf{Q}}e^{-i\mathbf{Q.R}_{n}}B_{Q}^{\dagger
}=\frac{1}{N_{s}}\sum_{n^{\prime }=1}^{N_{s}}B_{n^{\prime }}^{\dagger }\sum_{%
\mathbf{Q}}e^{i\mathbf{Q.(\mathbf{R}_{n^{\prime }}-\mathbf{R}_{n})}%
}=B_{n}^{\dagger }  \tag{6.12}  \label{6.12}
\end{equation}%
since the sum over $\mathbf{Q}$ is equal to 0 for $n^{\prime }\neq n$ and to 
$N_{s}$ for $n^{\prime }=n$.

These Frenkel excitons are expected to be composite bosons. This is easily
seen from their commutation rules. By using Eq. (5.3) for electron-hole
pairs on site $n$, we readily find%
\begin{equation}
\left[ B_{Q},B_{Q^{\prime }}\right] _{-}=0.  \tag{6.13}  \label{6.13}
\end{equation}

If we now turn to the other commutator, it reads%
\begin{equation}
\left[ B_{Q^{\prime }},B_{Q}^{\dagger }\right] _{-}=\frac{1}{N_{s}}%
\sum_{n^{\prime }=1}^{N_{s}}\sum_{n=1}^{N_{s}}e^{-i\mathbf{Q}^{\prime }.%
\mathbf{\mathbf{R}_{n^{\prime }}}}e^{i\mathbf{Q.\mathbf{R}_{n}}}\left[
B_{n^{\prime }},B_{n}^{\dagger }\right] _{-}.  \tag{6.14}  \label{6.14}
\end{equation}%
We then use Eq. (5.4) for the commutator $\left[ B_{n^{\prime
}},B_{n}^{\dagger }\right] _{-}$. The term in $\delta _{nn^{\prime }}$ gives 
$N_{s}^{-1}\sum_{n}e^{i(\mathbf{Q-Q}^{\prime }).\mathbf{\mathbf{R}_{n}}}$,
which is equal to 0 for $\mathbf{Q}\neq \mathbf{Q}^{\prime }$ and 1 for $%
\mathbf{Q}=\mathbf{Q}^{\prime }$; so that we end with\ 
\begin{equation}
\left[ B_{\mathbf{Q}^{\prime }},B_{\mathbf{Q}}^{\dagger }\right] _{-}=\delta
_{\mathbf{Q}^{\prime }\mathbf{Q}}-D_{\mathbf{Q}^{\prime }\mathbf{Q}} 
\tag{6.15}  \label{6.15}
\end{equation}%
where the deviation-from-boson operator $D_{\mathbf{Q}^{\prime }\mathbf{Q}}$
for Frenkel excitons, which comes from the deviation-from-boson operator for
electron-hole pairs $D_{n^{\prime }n}$ appearing in $\left[ B_{n^{\prime
}},B_{n}^{\dagger }\right] _{-}$ reads as 
\begin{equation}
D_{\mathbf{Q}^{\prime }\mathbf{Q}}=\frac{1}{N_{s}}\sum_{n=1}^{N_{s}}e^{i%
\mathbf{(Q-Q}^{\prime }\mathbf{).\mathbf{R}_{n}}}(a_{n}^{\dagger
}a_{n}+b_{n}^{\dagger }b_{n}).  \tag{6.16}  \label{6.16}
\end{equation}

We can note that, as for Wannier excitons, this deviation-from-boson
operator gives 0 when acting on the pair vacuum state%
\begin{equation}
D_{\mathbf{Q}^{\prime }\mathbf{Q}}\left\vert 0\right\rangle =0.  \tag{6.17}
\label{6.17}
\end{equation}

This leads us to conclude that, in order to describe the interactions
between Frenkel excitons properly, it is necessary to follow a path similar
to the one we have used for Wannier excitons, namely, to define the Pauli
scatterings of two Frenkel excitons for carrier exchanges without Coulomb
interaction and the Coulomb scatterings of two Frenkel excitons for carrier
interactions without carrier exchange. As for Wannier exciton, the composite
nature of the Frenkel excitons makes impossible a clean description of the
interactions between two excitons as a potential, the only well-defined
quantity again being the "creation potential" of the $Q$ exciton. The
calculation of these Pauli and Coulomb scatterings for Frenkel excitons,
necessary to handle their many-body physics properly, will be done in the
forthcoming publication.

\subsection{Interacting Frenkel exciton hamiltonian}

If we have more than one electron-hole pair, the Coulomb part $V_{coul}$ of
the Frenkel exciton hamiltonian, given in Eq. (4.23), is going to play a
role. This leads us to rewrite $H^{(F)}$ as $H^{(F)}=H_{X}$ with 
\begin{equation}
H_{X}=H_{X}^{(0)}+V_{coul},  \tag{6.18}  \label{6.18}
\end{equation}%
where $H_{X}^{(0)}$, as given in Eq. (6.2), corresponds to the pair
hamiltonian $H_{pair}$ plus the part of the Coulomb interaction $V_{trans}$
allowing for the excitation transfer.

The remaining part $V_{coul}$, given in Eq. (4.25), corresponds to all
direct Coulomb interactions between two electrons, two holes, and one
electron and one hole in different sites. This operator is going to generate
all many-body effects between excitons induced by Coulomb interactions. In
addition to them, as for Wannier excitons, also exist many-body effects
induced by Pauli exclusion through \ the fact that Frenkel excitons are not
elementary bosons. The scatterings associated to $V_{coul}$ and to the
deviation-from-boson operators will be calculated in a forthcoming
publication.

\section{Conclusion}

In this paper, we have derived the creation operator for Frenkel exciton
starting from the microscopic hamiltonian for free electrons in a periodic
lattice. Let us summarize the main steps of the derivation.

1) We first isolate one ion located on site $n$ and we introduce the atomic
states $\left\vert \nu n\right\rangle $, eigenstates for this particular
ion. They form a complete basis for one-electron states.

2) If we let $n$ running over all sites, the states $\left\vert
vn\right\rangle $ form an overcomplete set. However, if the states of
physical interest are the two lowest atomic levels, $\nu =0$ and $\nu =1$,
the states $\left\vert \nu n\right\rangle $ with $\nu =(0,1)$ are
essentially orthogonal in the tight binding limit, i.e., when the overlap of
the $\nu =(0,1)$ wave functions on different sites is negligible. This
allows us to use the $\left\vert \nu =(0,1)\text{ }n\right\rangle $ as a
one-electron \ basis to describe Frenkel exciton in second quantization.

3) The large energy difference between the atomic states $\nu =(0,1)$ leads
us, in the electron-electron interaction written in terms of the creation
operators $a_{\nu n}^{\dagger }$ for these $\left\vert \nu n\right\rangle $
states, to only keep the terms which conserve the number of electrons in
state $\nu =0$ and in state $\nu =1$ separately.

4) When written in terms of electron-hole pairs with $a_{n}^{\dagger
}=a_{1n}^{\dagger }$ and $b_{n}^{\dagger }=a_{0n}$, the electron-electron
interaction generates a constant term, which renormalizes the band gap. It
also generates one-body contributions in $b_{n}^{\dagger }b_{n}$ and $%
a_{n}^{\dagger }a_{n}$ which dress the $(\nu =0)$ and $(\nu =1)$ atomic
levels, the electron and hole energies differing from the atomic energies
due to the Coulomb interaction with a kind of jellium having one electron in
the ground state of each ion site.

5) Finally, the electron-electron interaction also generates an intersite
contribution $V_{inter}$ given in Eq. (4.22a) and an intrasite contribution $%
V_{intra}$ which insures local neutrality; so that the lowest excited states
have one electron and one hole on the \textit{same} site.

6) The corresponding degenerate subspace $a_{n}^{\dagger }b_{n}^{\dagger
}\left\vert 0\right\rangle =B_{n}^{\dagger }\left\vert 0\right\rangle $\
with $n$ running over all sites, is split by the part $V_{trans}$ given by
Eq. (4.21) of the intersite contribution $V_{inter}$ which allows to
transfer the excitation from one site to the other.

7) The resulting eigenstates correspond to a set of delocalized excitations,
known as Frenkel excitons. Their creation operators read 
\begin{equation*}
B_{Q}^{\dagger }=\frac{1}{\sqrt{N_{s}}}\sum_{n=1}^{N_{s}}e^{i\mathbf{Q.R}%
_{n}}B_{n}^{\dagger },
\end{equation*}%
where $N_{s}$ is the number of ion sites, these ions being located at $%
\mathbf{R}_{n}$ on a periodic lattice.

8) These Frenkel excitons are composite bosons. Their many-body effects thus
have to be handled along a procedure similar to the one we have used for
Wannier excitons. We are going to show that they predominantly interact
through the Pauli exclusion principle between their electron-hole components
which make them differing from elementary bosons and which produces "Pauli
scatterings" to describe carrier exchanges without carrier interaction.
Frenkel excitons also interact through the $V_{coul}$ part of the intersite
interaction given in Eq. (4.25), which contains electron-electron, hole-hole
and electron-hole direct processes, i.e., processes in which the carriers
stay on their site.

The composite-boson many-body theory appropriate to Frenkel excitons will be
presented in a forthcoming publication, the present work providing the
necessary tools to build this theory on solid grounds.

\acknowledgments

W. V. P. is supported by the Ministry of Education of France, the Russian
Science Support Foundation, and the President of Russia program for young
scientists.

\section{\textbf{Figure captions}}

\textbf{Fig 1. }Electron-electron potentials, given in Eq. (3.15).

(a) Terms in $a_{0n_{1}}^{\dagger }a_{0n_{2}}^{\dagger }a_{0n_{2}}a_{0n_{1}}$%
, in which the two electrons stay in their atomic ground state $\nu =0$.

(b) Terms in $a_{1n_{1}}^{\dagger }a_{1n_{2}}^{\dagger }a_{1n_{2}}a_{1n_{1}}$%
, in which the two electrons stay in their atomic excited state $\nu =1$.

(c) Terms in $a_{0n_{1}}^{\dagger }a_{1n_{2}}^{\dagger }a_{1n_{2}}a_{0n_{1}}$%
, in which one electron stays in the ground state, while the another one
stays in the excited state.

(d) Terms in $a_{1n_{1}}^{\dagger }a_{0n_{2}}^{\dagger }a_{1n_{2}}a_{0n_{1}}$%
, in which the electron on site $n_{1}$ is excited from the ground state $%
\nu =0$ to the excited state $\nu =1$, while the electron on site $n_{2}$
returns to its ground state.

In these four processes, the number of electrons in the ground state and in
the excited state are separately conserved.\ 

\textbf{Fig 2.} Excitation of an electron from the ground state on site $%
n_{1}$ to the excited state on site $n_{2}$: this corresponds to the
creation of an electron-hole pair on sites ($n_{2}$, $n_{1}$).

\textbf{Fig. 3}. (a) Electron-electron interaction $\widetilde{V}_{ee}$,
given in Eq. (4.6).

(b) Hole-hole interaction $\widetilde{V}_{hh}$, given in Eq. (4.13).

(c) Direct electron-hole interaction $V_{eh}^{(dir)}$, given in Eq. (4.8).

(d) Exchange electron-hole interaction $V_{eh}^{(exch)}$, given in Eq.
(4.10).

\textbf{Fig. 4}. Interaction between sites, described by $V_{inter}$, given
in Eq. (4.25).

(a) The part $V_{trans}$, given in Eq. (4.24), describes the destruction of
an electron-hole pair on site $n_{1}$ and its recreation on site $n_{2}$.\ 

(b) Direct interactions between two electrons, corresponding to $\widetilde{V%
}_{ee}$, given in Eq. (4.6), (c) between two holes, corresponding to $%
\widetilde{V}_{hh}$, given in Eq. (4.13), and (d) between one electron and
one hole, corresponding to $\widetilde{V}_{eh}^{(dir)}$, given in Eq. (4.23).

\end{document}